\documentclass[12pt,letterpaper]{article}
\pdfoutput=1

\usepackage{amsmath,amssymb,calc, amsthm,bbm, epsfig,psfrag, mathtools}
\usepackage{graphicx, enumerate}
\usepackage[numbers,sort&compress]{natbib}
\usepackage[dvipsnames]{xcolor}
\usepackage{physics}
\usepackage{dsfont}
\usepackage{tensor}
\usepackage{comment}

\usepackage[mathscr]{euscript}

\usepackage[utf8]{inputenc} 

\newcommand{\Comment}[1]{{}}
\definecolor{darkblue}{rgb}{0.15,0.35,0.55}
\definecolor{reddish}{rgb}{0.65, 0.2, 0.2}
\usepackage[linktocpage=true]{hyperref}
\hypersetup{
colorlinks=true,
citecolor=darkblue,
linkcolor=reddish,
urlcolor=darkblue,
pdfauthor={},
pdftitle={},
pdfsubject={}
}

\makeatletter
\renewcommand\section{\@startsection {section}{1}{\z@}%
                                   {-3.5ex \@plus -1ex \@minus -.2ex}%nn
                                   {2.3ex \@plus.2ex}%
                                   {\normalfont\large\bfseries}}
\renewcommand\subsection{\@startsection{subsection}{2}{\z@}%
                                     {-3.25ex\@plus -1ex \@minus -.2ex}%
                                     {1.5ex \@plus .2ex}%
                                     {\normalfont\bfseries}}
\makeatother

\let\non\nonumber

\def\AdS{{\rm AdS}}

\def\bea#1\eea{\begin{align}#1\end{align}}

\def\bes #1\ees{\begin{split}#1\end{split}}

\newcommand{\be}{\begin{equation}}
\newcommand{\ee}{\end{equation}}

\newcommand{\bma}{\begin{pmatrix}}
\newcommand{\ema}{\end{pmatrix}}

\newcommand{\half}{\frac{1}{2}}

\let\x=\xi

\let\S=\Sigma

\def\th{\theta}

\def\x{\xi}

\def\d{{\rm d}}

\newcommand{\C}[1]{$(\ref{#1})$}

\def\Tr{{\rm Tr}}

\newlength{\bredde}
\def\slash#1{\settowidth{\bredde}{$#1$}\ifmmode\,\raisebox{.15ex}{/}
\hspace*{-\bredde} #1\else$\,\raisebox{.15ex}{/}\hspace*{-\bredde}
#1$\fi}

\newsavebox{\zzzbar}
\sbox{\zzzbar}
  {\setlength{\unitlength}{0.9em}
  \begin{picture}(0.6,0.7)
  \thinlines
  \put(0,0){\line(1,0){0.6}}
  \put(0,0.75){\line(1,0){0.575}}
  \multiput(0,0)(0.0125,0.025){30}{\rule{0.3pt}{0.3pt}}
  \multiput(0.2,0)(0.0125,0.025){30}{\rule{0.3pt}{0.3pt}}
  \put(0,0.75){\line(0,-1){0.15}}
  \put(0.015,0.75){\line(0,-1){0.1}}
  \put(0.03,0.75){\line(0,-1){0.075}}
  \put(0.045,0.75){\line(0,-1){0.05}}
  \put(0.05,0.75){\line(0,-1){0.025}}
  \put(0.6,0){\line(0,1){0.15}}
  \put(0.585,0){\line(0,1){0.1}}
  \put(0.57,0){\line(0,1){0.075}}
  \put(0.555,0){\line(0,1){0.05}}
  \put(0.55,0){\line(0,1){0.025}}
  \end{picture}}

\newcommand{\eq}[1]{\begin{align}\begin{split}#1\end{split}\end{align}}

\newfont{\goth}{ygoth.tfm scaled 1200}                   % gothic font (usual)
 \numberwithin{equation}{section}

\def\1{{(1)}}
\def\2{{(2)}}
\def\3{{(3)}}

\newcommand{\overbar}[1]{\mkern 1.5mu\overline{\mkern-1.5mu#1\mkern-1.5mu}\mkern 1.5mu}

\def\bz{{\bar{z}}}

\def\be{ \begin{equation} }
\def\ee{ \end{equation}}

\def\cot{{\rm cot}}
\def\det{{\rm det}}

\def\exp{{\rm exp}}

\def\log{{\rm log}}

\def\half{\frac{1}{2}}

\def\one{{\hbox{ 1\kern-.8mm l}}}
\def\bz{{\bar{z}}}

\def\bT{{\bar{T}}}

\def\bz{{\bar{z}}}

\def\CH {{\cal H}}

\def\CL {{\cal L}}

\def\CO {{\cal O}}

\def\CO {{\cal O}}

\def\CH {{\cal H}}

\def\IH{\mathbb{H}}

\def\IR{{\mathbb{R}}}

\def\IZ{{\mathbb{Z}}}

\def\rmk#1{\bigskip\noindent{\bf Remark} }
\def\cnj#1{\bigskip\noindent{\bf Conjecture:} }
\def\TT{{T\overbar{T}}}
\def\CL{{\mathcal{L}}}

\def\x{{\mathcal X}}

\begin{document}
\begin{titlepage}

\begin{center}

{May 1, 2020}
\hfill         \phantom{xxx}  EFI--20-7

\vskip 2 cm {\Large \bf 
Defining the $\TT$ Deformation on $\AdS_2$}
\vskip 1.25 cm {\bf T. Daniel Brennan, Christian Ferko, Emil Martinec and Savdeep Sethi}\non\\
\vskip 0.2 cm
 {\it Enrico Fermi Institute \& Kadanoff Center for Theoretical Physics \\ University of Chicago, Chicago, IL 60637, USA}

\vskip 0.2 cm

\end{center}
\vskip 1.5 cm

\begin{abstract}
\baselineskip=18pt

\noindent We show that the $\TT$ deformation of two-dimensional quantum field theory on $\AdS_2$ is well-defined and solvable at the quantum level. Flow equations for the energy spectrum and partition function are derived in analogy with the flat space case. As a non-trivial check, we perturbatively compute the deformed energy spectrum for the case of a free scalar field.  We analyze the high energy density of states of the deformed theory and find a Hagedorn growth of states.  

\end{abstract}

\end{titlepage}

\tableofcontents

\section{Introduction} \label{sec:intro}
\subsection{Background}

The $\TT$ deformation of two-dimensional quantum field theory is very special because it is an irrelevant deformation that is 
solvable for flat spacetimes. 
Usually turning on an irrelevant operator is bad news for understanding the ultraviolet physics of a system. Turning on one such operator usually requires turning on an infinite number of such operators as one tries to reverse a renormalization group flow. An irrelevant deformation can change the ultraviolet definition of the theory and the fundamental high-energy degrees of freedom.

However, the case of $T\bar{T}$ is very special because this composite irrelevant operator is quantum mechanically well-defined, and one can track what happens to the energies of the system as a function of the deformation parameter. It is in this sense that the deformation is solvable: some quantities of interest in the deformed theory, like the finite-volume spectrum or $S$-matrix, can be computed in terms of the corresponding quantities in the undeformed theory~\cite{Zamolodchikov:2004ce, Smirnov:2016lqw, Cavaglia:2016oda, Dubovsky:2012wk}.

For this irrelevant operator, one can partially reverse the renormalization group flow and understand something about the ultraviolet physics. The result is surprising. The high-energy density of states on a cylinder exhibits a Hagedorn growth resembling that of a string theory, while the low-energy physics resembles conventional local quantum field theory. This suggests the existence of a new and still mysterious structure somewhere between a local quantum field theory and a full fledged string theory.  There is considerable excitement around this topic currently, and we refer the reader to the review~\cite{Jiang:2019hxb}.

All the original arguments for the $\TT$ deformation use a flat spacetime; typically either a cylinder or $\IR^{1,1}$. It is unclear if the deformed theory can even be defined on an arbitrary fixed curved manifold, or more ambitiously be consistently coupled to gravity. In this project, we will discuss the first question of two-dimensional quantum field theory on a manifold $\mathcal{M}$ with a fixed curved background metric $g$, deformed by an irrelevant operator. 

This approach should be contrasted with various holographic proposals for defining the $\TT$ deformation on curved spacetime, all of which involve some dynamical three-dimensional bulk metric and a two-dimensional boundary theory \cite{McGough:2016lol, Kraus:2018xrn, Taylor:2018xcy, Hartman:2018tkw, Mazenc:2019cfg, Hirano:2020nwq}. These interesting holographic proposals retain some of the solvability properties of $\TT$, at least in certain regimes. For instance for conformal field theory in the large $N$ or large central charge limit, one can attempt to compute the $\TT$-deformed partition function on a sphere using a cut-off $\AdS$ prescription \cite{Caputa:2019pam}. However, this approach applies for the sign of the deformation which is usually called the `bad sign' because most energies of the deformed system are complex. Interpreting the resulting deformed theory is an interesting challenge. By contrast, our approach in this work applies to either sign of the deformation parameter. Another interesting observation involving a dynamical metric, first studied in \cite{Dubovsky:2017cnj} and nicely generalized in \cite{Ishii:2019uwk, Okumura:2020dzb}, relates the $\TT$ deformation to certain perturbations of the metric and dilaton around vacuum solutions to 2D dilaton systems. Unlike these proposals, we do not use holography or dynamical gravity; rather, we will always work in two dimensions with a fixed background metric.

For completely general spacetimes $(\mathcal{M} , g )$, the usual arguments for the solvability of $\TT$ fail because the metric might not be translation-invariant, or have any isometries. These are key ingredients used in the usual analysis of the $\TT$ deformation. The only cases for which solvability might be straightforwardly preserved are theories which are almost entirely topological. For instance, one can study the $\TT$ deformation of two-dimensional pure Yang-Mills theory on any background metric \cite{Ireland:2019vvj}, since this theory has no propagating degrees of freedom and therefore almost no dependence on the metric. However, if one couples two-dimensional Yang-Mills to charged matter, the resulting theory is no longer topological; one can still study the $\TT$ deformation of this coupled theory in flat space \cite{Brennan:2019azg}, but the conclusions concerning pure $\mathrm{YM}_2$ with an arbitrary background metric will no longer hold.

Nonetheless, one might hope that certain highly symmetric spacetimes $(\mathcal{M} , g)$ might still admit solvable $\TT$-like deformations as in the flat space case. For instance, one might restrict to spaces of constant non-zero curvature. This question has been investigated in \cite{Jiang:2019tcq}, which proposed an interesting generalization of $\TT$ to spaces of constant curvature. 
The proposed prescription leads to a definition of $\TT$ which does not appear to enjoy the same factorization properties as the corresponding operator in flat space. In this paper, we will propose a different definition of $\TT$ in spaces of constant negative curvature. As we will discuss in Section \ref{sec:tt_in_ads} and Appendix \ref{app:2point}, this modified definition yields an operator that does factorize, which leads to similar solvability properties as in flat space.

The layout of this paper is as follows: in subsection \ref{sec:summary}, we summarize the main technical results of the paper. Section \ref{sec:tt_in_ads} proposes a new definition of the $\TT$ operator in spaces of constant negative curvature and investigates the properties of this operator. In Section \ref{sec:flow_spectrum}, we show that deformation by this operator leads to essentially the same differential equation for the spectrum as the usual $\TT$ deformation in flat space. Similarly, Section \ref{sec:flow_partition} shows that our prescription also leads to a flow equation for the partition function which is totally analogous to the flow equation in flat space. Finally, Appendices \ref{app:2point} and \ref{app:perturbative} collect some technical arguments regarding factorization of our $\TT$ operator and a perturbative check of the resulting flow equation for the energy levels, respectively.

\subsection{Summary}\label{sec:summary}

To summarize our results: we find that the $\TT$ operator can be unambiguously defined by point-splitting in a way which leads to a solvable deformation for theories on $\AdS_2$. We take the $\AdS_2$ metric to be,
\eq{\label{ads_metric}
ds^2_{\AdS_2}=\frac{a^2}{\sin^2\sigma}(-d\tau^2+d\sigma^2)~,
}
with $a$ of length dimension $1$ and the coordinates dimensionless. 
It is unclear whether the $\TT$ deformation is solvable on general curved manifolds. 
However, $\AdS_2$ is quite special. 
More precisely, $\AdS_2$ is the unique non-flat, homogeneous, non-compact %2D 
Lorentzian manifold $\mathcal{M}$ with a transitive global isometry group. 
Because of this, we can define the $\TT(x)$ operator as follows, 
\eq{ \label{defT}
\TT(x):=\half \lim_{y \to x}\left(I_{ac'}(x,y)I_{bd'}(x,y)-\eta_{ab}\eta_{c'd'}\right)T^{ab}(x)T^{c'd'}(y)~,
}
where $x,y$ are separated along a spatial (constant $\tau$) geodesic. 
The stress tensor appearing in~\C{defT}\ is referred to a flat frame in the tangent space, related to the usual stress tensor with coordinate indices using frame fields $e^a_{~\mu}$,
\eq{
T^{ab}(x)=e^{a}_{~\mu}e^b_{~\nu}T^{\mu\nu}(x)~,
}
with $(a,b),(c',d')$ the indices for the orthonormal frame bundle above the points $x,y$ respectively. Additionally, $I_{ab'}(x,y)$ is the parallel transport tensor of the frame bundle along the geodesic from $y\to x$ and $\eta_{ab}$ is the frame bundle metric. 

The fact that $x,y$ are separated along a spatial geodesic implies that the parallel transport tensor is trivial. Because of this, we find that the operator above factorizes 
\eq{
\langle \TT(x)\rangle=\half \lim_{y \to x}\left(I_{ac'}(x,y)I_{bd'}(x,y)-\eta_{ab}\eta_{c'd'}\right)\langle T^{ab}(x)\rangle\langle T^{c'd'}(y)\rangle~.
}
This result allows us to define a flow equation for the energy spectrum:
\eq{
\partial_\lambda E_n=-\frac{1}{2\pi }\left(E_n \partial_a E_n+\frac{P_n^2}{a}\right)~,
}
where $a$ is the $\AdS_2$ length scale and $\lambda$ is the $\TT$ deformation parameter. 
Somewhat surprisingly, we find that this result is of the same form as the analogous equation for the deformed energies on a flat cylinder where the radius $R$ has been exchanged with the $\AdS_2$ scale factor $a$.
One might have thought that the non-trivial curvature of $\AdS_2$, and exponential gravitational redshift, would significantly distort the spectrum; nevertheless it looks more or less the same as one would get on a flat cylinder.
This flow equation for the energies implies an analogous property of the thermal partition function in the zero-momentum sector:
% \emil{revise when section 4 is fixed!!!???}
\eq{ \partial_\lambda Z = {1\over 2\pi a} \left\{ \partial_\beta\partial_a - {\beta\over a}\partial_\beta^2 - {1\over \beta} \partial_a \right\} Z .
}
In the case where the undeformed theory is a CFT, we find that the deformed energies for states with $P_n=0$ are of the form 
\eq{
E_n=-\frac{\pi a}{\lambda}\left(1-\sqrt{1+\frac{2\lambda E_n^{(0)}}{\pi a}}\right)~,
}
where $E_n^{(0)}$ is the undeformed energy. We show that this spectrum agrees to leading order with the energy spectrum of the $\TT$-deformed free scalar field. 

As in the case of $\TT$-deformed theories on flat space, we find that the high energy density of states for $\TT$-deformed theories on $\AdS_2$ is enhanced. For undeformed theories with a Cardy-like high energy density of states, the deformed theories exhibit Hagedorn growth.  Note that this matches the expectation that a $\TT$-deformed theory on $\AdS_2$ is no longer a local quantum field theory, but should behave more like a string theory.

\section{The \texorpdfstring{$\TT$}{TT} Deformation in \texorpdfstring{$\AdS_2$}{AdS2}} \label{sec:tt_in_ads}

In the following discussion we will be concerned with quantum field theories on $\AdS_2$, which is the unique homogeneous, non-compact non-flat spacetime in two dimensions. This spacetime and its Euclidean version $\IH_2$ can be realized as quotients of either  $SL(2,\IR)$ or its universal cover $\widetilde{SL}(2,\IR)$ \cite{Kitaev:2017hnr}
\eq{
\AdS_2= \widetilde{SL}(2,\IR)/SO(1,1)~,
\quad\quad
\IH_2 = SL(2,\IR)/SO(2)
~.
}
These realizations preserve the left action by $SL(2,\IR)$ which thus acts faithfully and transitively as a global isometry group.  Because it is a homogeneous space, $\AdS_2$ has constant (negative) curvature; it can be realized as the infinite strip with metric
\eq{
\label{adsmetric}
ds^2=\frac{a^2}{\sin^2\sigma}(-d\tau^2+d\sigma^2)~,
}
where $\sigma\in (0,\pi)$, $\tau \in \IR$ and only $a$ is a dimensionful parameter.

Quantum field theory on anti-de Sitter space is slightly more unusual than on standard flat spacetime $\IR^{1,1}$.  Because it has constant negative curvature, volumes grow exponentially near its conformal boundaries $\sigma=0,\pi$. This implies for example that Gauss's law is exponentially suppressed as are propagators of massive fields. This property of $\AdS$ provides a nice IR regulator as discussed in \cite{Callan:1989em}. 

For our purposes, we will need the fact that two point functions factorize in $\AdS_2$ as the insertion points go to infinite separation. In fact, the exponential growth of volume implies that in some sense cluster decomposition occurs faster in $\AdS$ than in flat space as a function of geodesic separation.

\subsection{Review of \texorpdfstring{$\TT$}{TT}}\label{sec:tt_review}
The $\TT$ deformation was originally described in the classic paper by Zamolodchikov \cite{Zamolodchikov:2004ce}. There he considered a Euclidean QFT with the following properties:
\begin{enumerate}
    \item \emph{Local translation and rotation symmetry}: This property implies the existence of a stress energy tensor $T_{\mu\nu}$ with the properties
    \be
    \partial_\mu T^{\mu\nu}=0~,
    \ee
    which in the 2D parametrization $T_{zz}=T,~\bar{T}=T_{\bar{z}\bar{z}},~\Theta=T_{z\bar{z}}$ can be written
    \be
    \partial_{\bz} T(z)=\partial_z \Theta(z)~,\quad \quad \partial_z \bar{T}(z)=\partial_\bz \Theta(z)~. 
    \ee
    \item\label{global_trans} {\emph{Global translation symmetry}}: This property implies that any 1-point function is independent of position
    \be
    \langle \CO_i(z)\rangle=\langle \CO_i(0)\rangle~, 
    \ee
    and that 
    any 2-point function 
    \be
    \langle \CO_i(z)\CO_j(z')\rangle=G_{ij}(z-z')~,
    \ee
    is only a function of the distance between the insertion points. 
    \item {\emph{Clustering}}: That there exists some direction of infinite length such that 
    \be
    \lim_{x\to \infty}\langle \CO_i(x)\CO_j(0)\rangle=\langle \CO_i\rangle\langle \CO_j\rangle~. 
    \ee
    \item {\emph{UV CFT}}: That the QFT in question is described by a CFT at short distances. 
\end{enumerate}
These conditions then require that we consider a theory on the flat plane or cylinder. Using these assumptions one can show that the operator 
\be
\TT:=\lim_{z'\to z}\left(T(z')\bar{T}(z)-\Theta(z')\Theta(z)\right)~,
\ee
defines a local operator. 

First, note that the conservation of the stress tensor implies 
\begin{align}\begin{split}\label{deriv}
\partial_\bz (T(z)\bar{T}(z')-\Theta(z)\Theta(z'))=(\partial_z+\partial_{z'})\Theta(z)\bar{T}(z')-(\partial_\bz+\partial_{\bz'})\Theta(z)\Theta(z')~,\\
\partial_z(T(z)\bar{T}(z')-\Theta(z)\Theta(z'))=(\partial_z+\partial_{z'})T(z)\bT(z')-(\partial_\bz+\partial_{\bz'})T(z)\Theta(z')~.
\end{split}\end{align}
Now using the operator product expansions
\begin{align}\begin{split}
    T(z)\Theta(z')=\sum_i A_i(z-z')\CO_i(z')~,\quad\quad\Theta(z)\Theta(z')=\sum_i C_i(z-z')\CO_i(z')~,\\
    \Theta(z)\bT(z')=\sum_i B_i(z-z')\CO_i(z')~,\quad\quad T(z)\bT(z')=\sum_i D_i(z-z')\CO_i(z')~,
\end{split}\end{align}
the equations \eqref{deriv} imply 
\begin{align}\begin{split}
&\sum_i \partial_\bz F_i(z-z')\CO_i(z')=\sum_i \Big(B_i(z-z') \partial_{z'}\CO_i-C_i (z-z')\partial_{\bz'}\CO_i\Big)~,\\
&\sum_i \partial_z F_i (z-z')\CO_i(z')=\sum_i\Big(D_i(z-z') \partial_{z'}\CO_i(z')-A_i(z-z')\partial_{\bz'}\CO_i(z')\Big)~, 
\end{split}
\end{align}
where 
\be
F_i(z-z')=D_i(z-z')-C_i(z-z')~. 
\ee
This implies that any operator arising in the OPE
\be
T(z)\bT(z')-\Theta(z)\Theta(z')=\sum_i F_i(z-z')\CO_i(z')~,
\ee
must either have a coordinate independent coefficient function $F_i(z-z')$ or is itself the derivative of another local operator:
\be
T(z)\bar{T}(z')-\Theta(z)\Theta(z')=\CO_{\TT}(z')+{\rm derivative~ terms}~.
\ee
This allows us to define the composite operator 
\be
\TT(z):=\CO_{\TT}(z)~. 
\ee
Note that  we have only defined $\TT$ up to derivative terms, but these contribute trivially  to one-point functions. 

\subsection{\texorpdfstring{$\TT$}{TT} in \texorpdfstring{$\AdS_2$}{AdS2}}
\label{sec:tt_in_ads2}

Now we will consider a mild generalization of Zamolodchikov's arguments on the properties of the $\TT$ deformation. Let us consider the case where assumption (\ref{global_trans}) is loosened to:
\begin{enumerate}
\setcounter{enumi}{1}
    \item {\emph{Transitive Global Isometry}}: Here we assume that $\forall x,y \in X$, where $X$ is a 2D Euclidean spacetime, there exists a global isometry $g\in \textit{Iso}(X)$ such that 
    \be
    g\cdot x=y~. 
    \ee
    Again, this implies that the expectation value of a one-point function is covariantly constant. 
    Additionally, it means that 2-point functions will be functions of the geodesic distance between the operators
    \be
    \langle \CO_i(x)\CO_j(y)\rangle=G_{ij}(d(x,y))~. 
    \ee
\end{enumerate}
These conditions allow us to more generally consider non-compact, homogeneous 2D spaces with a transitive isometry group. As it turns out, there is a unique nonflat manifold satisfying these conditions:  $\AdS_2$, which we will consider in either Lorentzian or Euclidean signature as appropriate. 

Now let us try to define the local operator $\CO_\TT$ in $\AdS_2$. As one might expect, there are many subtleties associated with this definition. First, the classical stress tensor $T_{\mu\nu}(x)$ is not a matrix, but rather resides in the tensor algebra of the tangent space over the point $x$. This means that the definition of the determinant requires extra consideration. In particular, one cannot simply contract tensor indices of products of stress tensors that are evaluated at different points by using the metric or epsilon symbol. 

There are now two ways to deal with this ambiguity to define \emph{the} $\TT$ operator. One suggestion, investigated in \cite{Jiang:2019tcq}, employs gravitational Wilson lines/parallel transport operators to define $\TT$ via point splitting
\be \label{jiang_ttbar_def}
\TT(y)=\half\lim_{x\to y}\left(I_{\mu \alpha'}I_{\nu \beta'}-g_{\mu\nu}g_{\alpha'\beta'}\right)T^{\mu\nu}(x)T^{\alpha'\beta'}(y)~,
\ee
where $\mu,\nu$ and $\alpha',\beta'$ are tensor indices for $T_x M$ and $T_y M$ respectively and
\be
I_{\mu \alpha'}={\rm exp}\left\{\int_\gamma ds_\nu \Gamma^\nu\right\}_{\mu\alpha'}\in \textit{Hom}(T_xM,T_yM)~,
\ee
where $\gamma$ is the minimal geodesic from $x$ to $y$.

In the coincident point limit, the operator (\ref{jiang_ttbar_def}) still defines a local $\TT$ operator despite the additional contribution from the gravitational Wilson line, as shown in \cite{Jiang:2019tcq}. However, the inclusion of the Wilson line spoils the factorization property which ensures that the expectation value of $\TT$ splits into a product of one-point functions. The effect of a non-trivial $I_{\mu \alpha'}$ is to introduce additional non-universal (i.e. theory-dependent) terms into $\langle \TT \rangle$ which depend on the structure of stress tensor two-point functions. In certain limiting cases -- such as the limit of very weak curvature, or in the limit where we $\TT$ deform a CFT with large central charge -- these non-universal terms are suppressed and approximate factorization is restored. In the general case, however, the obstruction to factorization is non-negligible and this definition of $\TT$ no longer admits simple solutions for quantities like the spectrum in the deformed theory.

To avoid this failure of factorization, we propose a different $\TT$ operator in $\AdS_2$. We will define $\TT$ to be
\eq{\label{TToperator}
\TT(y)&=\lim_{x\to y} C(x,y):=\lim_{x\to y} (I^{ac'}I^{bd'}-g^{ab}g^{c'd'})T_{ab}(x)T_{c'd'}(y)~,\\
&T_{ab}(x)=e_{a\mu}(x) e_{b\nu}(x)T^{\mu\nu}(x)~,
}
where we are taking the limit along a \emph{spatial geodesic}, $e_{a\mu}(x)$ are frame fields and $I^{ab'}$ is the parallel transport of the frame bundle:
\eq{
I^{ab'}(x,y)={\rm exp}\left\{\int_{\gamma(x,y)} \omega_\mu ds^\mu\right\}^{ab'}~, 
}
where $\omega_\mu^{ab}$ is the spin connection and $\gamma(x,y)$ is a geodesic  from $x\to y$. The components of the spin connection can be found in~\C{spinconnection}. 

This definition is preferable because in $\AdS_2$ the parallel transport along a spatial geodesic (in our conventions a path along the $\sigma$-direction)
is trivial on the frame bundle:%
\footnote{For any two spacelike separated points in $\AdS_2$, one can always use a boost isometry to go to a coordinate frame where the two points lie on a surface of constant $\tau$.} 
\eq{
I(\gamma_\sigma)=\mathds{1}~.
}
Since we are taking the coincident limit, it is clear that the local properties that allow the existence of $\TT(x)$ in flat space also apply to $\AdS_2$. We note that in any flat space the spin connection vanishes identically, and we could therefore define $\TT$ using this approach in any such space, like the examples studied in~\cite{Cardy:2018sdv}. 

Now from the transitivity of $\AdS_2$, we find that the Lorentz scalar $C(x,y)$ can only depend on the geodesic distance between $x,y$:
\eq{
C(x,y)=C\big(d(x,y)\big)~.
}
Following the analysis of \cite{Jiang:2019tcq}, we see that conservation of the stress tensor implies that the 2-point function is actually independent of geodesic distance between $x,y$:
\eq{
C(x,y)=C_0~.
}
See Appendix \ref{app:2point} for a detailed computation.

Since the 2-point function is coordinate independent, we can relate its expectation value at close points to its value  at asymptotically distant points where the 2-point function clusters. This result implies that the $\TT$ operator splits as a product of 1-point functions:
\eq{
(I^{ac'}I^{bd'}-g^{ab}g^{c'd'})\langle T_{ab}(x)T_{c'd'}(y)\rangle
&=(I^{ac'}I^{bd'}-g^{ab}g^{c'd'})\langle T_{ab}(x)\rangle \langle T_{c'd'}(y)\rangle
~,\\
&=(I^{ac'}I^{bd'}-g^{ab}g^{c'd'})\langle T_{ab}(0)\rangle \langle T_{c'd'}(0)\rangle~.
}
\noindent Therefore,   the composite $\TT(x)$ operator \eqref{TToperator} is both well defined on $\AdS_2$ 
and  acts diagonally on energy eigenstates. 

In curved spaces, one must account for the effects of the Weyl anomaly.  In a metric of the form
\eq{
ds^2 = g_{\mu\nu}\,dx^\mu dx^\nu = -e^{2\varphi} dx^+dx^-~, \qquad\quad
x^\pm=\tau\pm\sigma~,
}
the expectation value of the stress tensor gains a contribution from the Weyl anomaly
\eq{
\label{weylanom}
\big\langle T_{\mu\nu} [g] \big\rangle &= \big\langle T_{\mu\nu} [\eta] \big\rangle
+ \theta_{\mu\nu}~, 
}
where $\eta$ is the flat metric and
\eq{
\label{Lioustress}
\theta_{++} &= \frac{c}{12\pi}e^\varphi\partial_+^2 e^{-\varphi}~,\qquad\qquad
\theta_{--} = \frac{c}{12\pi}e^\varphi\partial_-^2 e^{-\varphi}~,
\\[.2cm]
\theta_{+-} &= \Bigl(\frac{c}{48\pi}R +\mu_0\Bigr) g_{+-}~, \qquad
R = 4 e^{-2\varphi}\partial_+\partial_- \varphi~.
}
The parameter $\mu_0$ arises from the introduction of a cosmological term $\mu_0 \sqrt{-g}$ in the action.  For the $\AdS_2$ metric~\eqref{adsmetric}, we have $\theta_{++}=\theta_{--}=-\frac{1}{48\pi}$, and $R=-1/a^2$.  The values of $\theta_{++}$ and $\theta_{--}$ represent the usual Casimir energy, arising here for the shift between Poincar\'e and global $\AdS_2$ coordinates \cite{Spradlin:1999bn}.  The contribution of $\theta_{+-}$ yields an infrared divergence in the energy integral due to the infinite spatial volume of $\AdS_2$, which we can cancel off by adjusting the cosmological constant $\mu_0$.  In what follows, we will assume that such an adjustment has been made.

%%%%%%%%%%%%%%%%%%%%%%%%%%%%%%%%%%%%%%%%%%%%
%%%%%%%%%%%%%%%%%%%%%%%%%%%%%%%%%%%%%%%%%%%%

\subsection{Deformed Lagrangian}

In this section, we will review the solution of the $\TT$ flow equation for the deformed Lagrangian $\mathcal{L} ( \lambda )$ of a free scalar field $\phi$ on $\AdS_2$. We stress that this is a purely classical result, unrelated to the preceding argument that the $\TT$ operator is well-defined by point-splitting. Indeed, the explicit solution for the deformed Lagrangian of scalars coupled to an arbitrary background metric was already written down in \cite{Bonelli:2018kik}, which follows from the analysis in \cite{Cavaglia:2016oda}. In addition, we will in this subsection ignore the effects of the trace anomaly.

Consider a general $\lambda$-dependent Lagrangian for a real scalar $\phi$ coupled to a background metric $g_{\mu \nu}$. For simplicity, we assume that the Lagrangian reduces to the usual free kinetic Lagrangian for $\phi$ at $\lambda = 0$:
\begin{align}
    \mathcal{L} ( \lambda = 0 ) = \frac{1}{2} g^{\mu \nu} \partial_\mu \phi \partial_\nu \phi .
    \label{initial_condition}
\end{align}
The $\TT$ flow will not introduce dependence on the undifferentiated field $\phi$, as one would have in a potential energy term, so the finite-$\lambda$ Lagrangian can only depend on the scalar quantity $g^{\mu \nu} \partial_\mu \phi \partial_\nu \phi$ and on the parameter $\lambda$. To ease notation, we define $\x = g^{\mu \nu} \partial_\mu \phi \partial_\nu \phi$ so that
\begin{align}
    \mathcal{L} ( \lambda ) = f ( \lambda, g^{\mu \nu} \partial_\mu \phi \partial_\nu \phi ) \equiv f ( \lambda, \x ) .
\end{align}
We now compute the components of the stress tensor,
\begin{align}
\label{Tdef}
    T_{\mu \nu}^{(\lambda)} = - \frac{2}{\sqrt{-g}} \frac{\delta S^{(\lambda)}}{\delta g^{\mu \nu}} , 
\end{align}
where $S^{(\lambda)}$ is the effective action of the deformed theory
\begin{align}
    S^{(\lambda)} = \int \, d^2 x \, \sqrt{-g} \, \mathcal{L} ( \lambda ) .
\end{align}
Taking the variation, one finds
\begin{align}
    T_{\mu \nu}^{(\lambda)} = g_{\mu \nu} f - 2 \frac{\partial f}{\partial \x} \partial_\mu \phi \partial_\nu \phi 
    \label{stress_tensor_components}
\end{align}
As discussed in Section \ref{sec:tt_in_ads2}, we define our $\TT$ definition using the stress tensor with frame bundle indices, namely
\begin{align}
    T_{ab} ( x ) = e_{a \mu} ( x ) e_{b \nu} ( x )  T^{\mu \nu} ( x ) .
    \label{TT_flat_indices}
\end{align}
In terms of $T_{ab}$, the flow equation for the Lagrangian is simply
\begin{align}
    \frac{\partial \mathcal{L}}{\partial \lambda} = \det ( T_{ab} ) .
\end{align}
The determinant of (\ref{TT_flat_indices}) is
\begin{align}
    \det ( T_{ab} ) &= \frac{1}{2} \left( \left( \tensor{T}{^a_a} \right)^2 - T^{ab} T_{ab} \right) \nonumber \\
    &= \frac{1}{2} \left( \left( e^{a \mu} e_{a \nu} \tensor{T}{^\nu_\mu} \right)^2 - e^{a \rho} e^{b \sigma} e_{a \mu} e_{b \nu} T^{\mu \nu} T_{\rho \sigma} \right) \nonumber \\
    &= \frac{1}{2} \left( \left( g^{\mu \nu} \tensor{T}{_\mu_\nu} \right)^2 - g^{\mu \rho} g^{\nu \sigma} T_{\mu \nu} T_{\rho \sigma} \right) .
    \label{det_T_expression}
\end{align}
That is, the determinant of our frame-bundle stress tensor $T_{ab}$ matches that of the usual stress tensor $T_{\mu \nu}$ when both are evaluated at a fixed point $x$. Although using frame bundle indices does not change the form of the expression (\ref{det_T_expression}) after taking the coincident point limit, we note that it was necessary to avoid parallel-transport contributions when defining the operator by point-splitting in Section \ref{sec:tt_in_ads2}. This distinction was necessary only to ensure factorization of two-point functions, but does not affect the analysis of the flow equation for the Lagrangian.

Evaluating (\ref{det_T_expression}) with the components (\ref{stress_tensor_components}), one finds
\begin{align}
    \det \left( T_{ab} \right) = f^2 - 2 f \x \frac{\partial f}{\partial \x} .
\end{align}
Thus the differential equation for the deformed Lagrangian becomes
\begin{align}
    \frac{d f}{d \lambda} = f^2 - 2 f \x \frac{\partial f}{\partial \x} , 
\end{align}
whose solution is (after replacing $\x = g^{\mu \nu} \partial_\mu \phi \partial_\nu \phi$ and imposing the initial condition (\ref{initial_condition}))
\begin{align}
    \mathcal{L} ( \lambda ) = \frac{1}{2 \lambda} \left( \sqrt{1 + 2 \lambda g^{\mu \nu} \partial_\mu \phi \partial_\nu \phi } - 1 \right) ~.
    \label{deformed_lagrangian_with_metric}
\end{align}
Expanding about $\lambda=0$, one finds
\begin{align}
    \mathcal{L} ( \lambda ) = \frac{1}{2} g^{\mu \nu} \partial_\mu \phi \partial_\nu \phi - \frac{\lambda}{4} \left( g^{\mu \nu} \partial_\mu \phi \partial_\nu \phi \right)^2 + \mathcal{O} ( \lambda ^2 ) ~.
    \label{leading_deformed_lagrangian}
\end{align}
We emphasize that this is a purely classical result that is true for any conformally flat metric 
$g_{\mu \nu}$. One could try to extend this analysis to the quantum effective action, see~\cite{Rosenhaus:2019utc} for a discussion of the one-loop renormalized effective action in flat space. Incorporating the Weyl anomaly, however, complicates the analysis. One might imagine taking it into account via a Liouville-type contribution to the effective action; we leave such a treatment to future work.
What is special about the $\AdS_2$ case is that one can also make statements about quantities in the quantum theory analogous to those in the flat space case, such as a flow equation for the spectrum, which we turn to next.

%%%%%%%%%%%%%%%%%%%%%%%%%%%%%%%%%%%%%%
%%%%%%%%%%%%%%%%%%%%%%%%%%%%%%%%%%%%%%

\section{Flow Equation for the Spectrum} \label{sec:flow_spectrum}
\subsection{Some Generalities}\label{generalities}

Now we can consider what happens to a theory on $\AdS_2$ when we deform it by the operator $\TT \equiv \det ( T_{ab} )$  defined in (\ref{det_T_expression}):
\be
S\mapsto S+\lambda \int d^2x\sqrt{g}\, \det ( T_{ab} ) ~.
\ee
In this case, the partition function by definition obeys the flow equation
\be\label{partflow}
\partial_\lambda \log Z  = \int d^2x\sqrt{g} \left\langle \det ( T_{ab} ) \right\rangle~,
\ee
which we can trivially rewrite as 
\be\label{flowtrivial}
\langle \partial_\lambda S\rangle= \int     d^2x\sqrt{g}\left\langle \det ( T_{ab} ) \right\rangle~.
\ee
As defined, this is a deformation of the Lorentzian or Euclidean path-integral. To find a flow equation for the energy spectrum, let us consider the thermal partition function defined as the Euclidean path-integral with periodic time $\tau\in [0,\beta)$, 
\be\label{path-integral}
Z[\beta] = \int \mathscr{D} \phi \, e^{-S_{E}}~, 
\ee
where $\mathscr{D} \phi$ schematically  denotes the path-integral over all fields. From the definition of the energy, 
\be\label{hamiltonian_from_Z}
\langle H \rangle = - {d\over d\beta} \log \, Z[\beta]~.
\ee
Differentiating the relation (\ref{hamiltonian_from_Z}) with respect to $\lambda$, we find
\begin{align}
    \label{hamiltonian_flow}
    \langle \partial_\lambda H \rangle &= - \partial_\beta \langle \partial_\lambda S_E \rangle = - \partial_\beta\int d^2x \, \sqrt{g} \, \langle \det ( T_{ab} ) \rangle ~, 
\end{align}
where we have used (\ref{flowtrivial}). 

Restricting (\ref{hamiltonian_flow}) to an energy eigenstate so that the expectation value of $\det(T)$ is time-independent allows us to evaluate the time integral. We then write the expectation value of the Hamiltonian as an integral of the Hamiltonian density $\mathcal{H}$ over a spatial slice $\Sigma$ to arrive at the flow equation
\begin{align}\label{energy_flow_start}
    \partial_\lambda \int_\Sigma dx \sqrt{g} \, \langle n \vert \, \mathcal{H} \, \vert  n \rangle = - \int_{\Sigma} dx \, \sqrt{g} \, \langle n | \, \det ( T_{ab} ) \, | n \rangle~,
\end{align}
where $\sqrt{g}$ refers to the full spacetime metric rather than the induced metric on a spatial slice. 
Equation (\ref{energy_flow_start}) is the starting point for the analysis of the flow equation for the energy levels which we will explore in the following subsections.

For our subsequent discussion, it will be convenient to derive some Ward identities in a setting general enough to accommodate $\AdS_2$. Let us assume a time-independent diagonal two-dimensional Euclidean metric of the form, 
\be
ds^2 = g_{tt} dt^2 + R^2 g_{\th\th}d\th^2,
\ee
where $t$ and $R$ have length dimension $1$ while $\th$ is dimensionless. Imagine sending the metric $g \rightarrow (1+\epsilon)g$ with $\epsilon$ constant. Since $\delta g_{tt} = \epsilon g_{tt}$, and $\beta$ is defined as the length $\int_{S^1} \sqrt{g_{tt}} \, dt$ of the time circle, this has the effect of shifting $\beta \to \left( 1 + \frac{1}{2} \epsilon \right) \beta$. Therefore, under this transformation the thermal partition function shifts as
\eq{
\partial_\epsilon  Z  &= \left( \frac{1}{2} \beta \partial_\beta + \frac{1}{2} R \partial_R \right)  \sum_n e^{-\beta E_n(R)}~,\\
& = \frac{1}{2} \sum_n \left( -\beta E_n - \beta  R {\partial E_n \over \partial R }  \right) e^{-\beta E_n(R)}~, 
}
which implies
\eq{\label{deriv_log_Z}
\partial_\epsilon  \log \, Z = - \frac{\beta}{2} \, \Big\langle E + R {\partial E \over \partial R } \Big\rangle.
}
Now we want to relate the left hand side of~\C{deriv_log_Z}\ to the stress tensor using the path-integral definition~\C{path-integral}. The conformal Ward identity gives
\be\label{Z_and_trace_T}
\partial_\epsilon  \log \, Z = \frac{1}{2} \left\langle \int d^2x \sqrt{g} \, \Tr(T) \right\rangle~.
\ee
Equating (\ref{deriv_log_Z}) and (\ref{Z_and_trace_T}), we find
\begin{align}\label{trace_T_to_expectation_values}
    \left\langle \int d^2 x \, \sqrt{g} \, \left( g^{tt} T_{tt} + \frac{1}{R^2} g^{\theta \theta} T_{\theta \theta} \right) \right\rangle = - \beta \left\langle E + R \frac{\partial E}{\partial R} \right\rangle.
\end{align}

Pick a spatial slice $\Sigma$ of constant $t$ to quantize along. With this choice of Cauchy surface, we can define the vectors
\be
\xi^\mu_t = \left( \frac{\partial}{\partial t} \right)^\mu ~, \quad \quad \hat{n}^\mu = \sqrt{g^{tt}} \xi_t^\mu~,
\ee
where $\hat{n}^\mu$ is the unit normal to $\S$ and $\xi_t^\mu$ is a time-like Killing vector.
Choosing an observer whose timelike Killing vector field is described by $\xi^\mu_t$, we define the energy on $\Sigma$ by 
\eq{\label{energy_def}
E&= - \int_\Sigma R \, d \theta \, \sqrt{g_{\theta \theta} } \,\hat{n}^\mu \xi^\nu_t T_{\mu\nu} ~.
}
Now the first term on the left side of (\ref{trace_T_to_expectation_values}) is
\begin{align}
    \int d t \, \int d \theta \, R \sqrt{g_{\theta \theta}} \, \xi^\mu_t\, \hat{n}^\mu \, \langle T_{\mu \nu} \rangle = - \int dt \, \langle E \rangle = - \beta \langle E \rangle ~ .
\end{align}
This cancels the $- \beta \langle E \rangle$ term on the right side of (\ref{trace_T_to_expectation_values}), leaving
\begin{align}\label{T_pressure}
    \int d \theta \, \frac{1}{R} \, \sqrt{\frac{g_{tt}}{g_{\theta \theta}}} \left\langle T_{\theta \theta} \right\rangle = - \left\langle R \frac{\partial E}{\partial R} \right\rangle ~ .
\end{align}
In the case of a flat cylinder, $g_{\theta \theta} = 1$ and $\theta \in [ 0 , 2 \pi ) $, this gives
\begin{align}
    \left\langle T_{\theta \theta} \right\rangle= - \frac{R^2}{2 \pi} \left\langle \frac{\partial E}{\partial R} \right\rangle~, 
\end{align}
which is the expected relation between the spatial components of the stress tensor and the pressure used in~\cite{Zamolodchikov:2004ce}.

\subsection{Flow Equation on the Conformal Cylinder}

Much of the subtlety of understanding the $\TT$ deformation on $\AdS_2$ reduces to understanding the role of varying the $\AdS$ length scale $a$ in the energy levels. As a warm up case, let us  first consider the $\TT$ deformation on a ``conformal cylinder" where we allow the overall scale $a$ to vary:
\be\label{conformalcylinder}
ds^2=a^2 (dt^2+R^2 d\theta^2)~.
\ee
Here  $\theta \in [0,2\pi)$ and the parameter $a$ are dimensionless, $t$ and $R$ have length dimension $1$.

For the metric~\C{conformalcylinder} and Cauchy surface $\Sigma$ of constant $t$, we can define 
\be
\hat{n}^\mu  =\frac{1}{a} \left(\frac{\partial}{\partial t}\right)^\mu~,\quad \quad \xi^\mu_t=\left(\frac{\partial}{\partial t}\right)^\mu~,
\ee
where $\hat{n}^\mu$ is the unit normal and $\xi_t^\mu$ is a time-like Killing vector. 
In the observer frame defined by the timelike Killing vector field  $\xi^\mu_t$, we  define the energy on $\Sigma$ by 
\eq{\label{E_cylinder_integral}
E&= - \int_\Sigma dx\sqrt{g_\Sigma} \,\hat{n}^\mu \xi^\nu_t T_{\mu\nu}= - R \int d\theta\, T_{tt}~.
}
In the quantum theory, this equates to the operator relation
\eq{\label{energyrelate}
\langle n|T_{tt} |n\rangle= - \frac{E_n}{2\pi R} ~ . % = - \frac{aE_n}{2\pi a R}~.
}
Since we work on the flat cylinder in this section, the usual assumptions for $\TT$ in flat space are assumed to hold. In particular, assumption (\ref{global_trans}) of Section \ref{sec:tt_review} posits that one-point functions are independent of position, which means that the integral over $\theta$ in (\ref{E_cylinder_integral}) is trivial.

Now let us consider the flow equation  (\ref{energy_flow_start}) for the $\TT$-deformed energy levels in the zero-momentum state,
\eq{
{\partial E_n \over \partial \lambda} = \partial_\lambda \int_\Sigma d\theta \sqrt{g} \,\langle n| \CH|n\rangle= - \frac{1}{a^4 R^2} \int_\Sigma d\theta \sqrt{g} \,\langle n|T_{tt} |n\rangle\langle n| T_{\theta\theta}|n\rangle~,
}
where we have used factorization of $\TT$.
Now let us consider the right hand side of this flow equation. Using \eqref{energyrelate}, we can rewrite the right hand side of the flow equation as
\eq{
 \frac{1}{a^4 R^2} \cdot \frac{E_n}{2\pi R}\int_0^{2\pi}  d\theta \, a^2 R \langle n|T_{\theta\theta} |n\rangle~.
}
Using~\C{T_pressure}\ in an energy eigenstate gives, 
\begin{align}
    \langle n| T_{\theta\theta}|n\rangle= - \frac{R^2}{2 \pi} %\Big\langle n \Big\vert \, 
    \frac{\partial E_n}{\partial R}~. %\, \Big\vert n \Big\rangle~. 
\end{align}
We therefore arrive at the flow equation with $P_n = 0$, 
\be
\label{cylinderburger}
{\partial E_n \over \partial \lambda} =- {1\over  2\pi a^2} E_n  \frac{\partial E_n}{\partial R}. 
\ee
This differs from the usual inviscid Burgers' equation by a factor of ${1\over a^2}$. We can check that this factor makes sense physically in two ways: scaling the spatial metric $g_{\th\th}$ by $a^2$ is tantamount to replacing $R \rightarrow a R$, which accounts for one factor of ${1\over a}$. On the other hand, rescaling the time metric, $g_{tt}$, by $a^2$ is tantamount to replacing $E_n \rightarrow {E_n \over a}$, which accounts for the remaining  factor of ${1\over a}$. 
Alternatively, we can regard $a$ as dimensionful and the coordinates together with $R$ as dimensionless.  From this perspective, the coupling $\lambda$ has dimensions of length squared, and so both sides of the equation scale as inverse length squared.

\subsection{Flow Equation on \texorpdfstring{$\AdS_2$}{AdS2}}
\label{sec:FlowAds2}

Now let us return to the discussion of the $\TT$ flow equation on $\AdS_2$. 
In order to evaluate the flow equation \eqref{energy_flow_start}, let us consider the theory on global Euclidean $\AdS_2$ which has metric 
\eq{
ds^2=\frac{a^2}{\sin^2\sigma}\left( d\tau^2+d\sigma^2\right)~,
}
where $\sigma\in (0,\pi)$. In these conventions, $\tau$ and $\sigma$ are dimensionless while the parameter $a$ has length dimension $1$. We can also consider thermal $\AdS_2$ with
$\tau \in [0,\beta)$ periodically identified.

To define the energy, we use the covariant definition
\eq{\label{E_AdS}
E&= - \int_\Sigma d \sigma \, \sqrt{g_{\sigma \sigma} } \,\hat{n}^\mu \xi^\nu_\tau T_{\mu\nu} ~,
}
together with a choice of observer whose trajectory in spacetime is specified by a time-like Killing vector $\xi^\mu_\tau$.  
We will pick the unit normal and time-like Killing vector as follows,  
\eq{
\hat{n}^\mu=\frac{\sin(\sigma)}{a}\left(\frac{\partial}{\partial \tau}\right)^\mu~,\quad \quad \xi^\mu_\tau =\frac{1}{a}\left(\frac{\partial}{\partial \tau}\right)^\mu~.
}
We choose these conventions so that the energy has mass dimension $1$. 
Now the definition of the energy implies
\eq{
E_n&= - \int_0^\pi d\sigma \frac{a}{\sin\sigma}\frac{\sin\sigma}{a}\frac{1}{a}\langle n|T_{\tau\tau}|n\rangle=
 - \frac{\pi}{a}\langle n|T_{\tau\tau}|n\rangle~.
}
Recall that we have chosen the constant $\mu_0$ in~\eqref{Lioustress} to cancel off the contribution of the Weyl anomaly to the expectation value of the stress tensor, so that the energy integral is infrared finite.

Now let us derive the flow equation starting from~\C{energy_flow_start}, 
\eq{\label{almostflow}
{\partial E_n \over \partial \lambda} &= - {1\over a}\partial_{\beta}\int d^2x \, \sqrt{g} \, \langle n| \det ( T_{ab} ) |n \rangle ~, \\
& = - {1\over a}\partial_{\beta}\int d^2x \sqrt{g} \, g^{\tau \tau} g^{\sigma \sigma} \left(\langle n|T_{\tau\tau} |n\rangle\langle n|T_{\sigma\sigma} |n\rangle-\langle n| T_{\tau\sigma}|n\rangle^2\right)~, \\
& =  {1\over a} \int d\sigma \sqrt{g} \, g^{\tau \tau} g^{\sigma \sigma} \left( {a E_n\over \pi} \langle n|T_{\sigma\sigma} |n\rangle+\langle n| T_{\tau\sigma}|n\rangle^2\right)~. 
}
The reasoning in section~\ref{generalities} leads to the relation 
\be
{1\over 2} \left\langle \int d^2 x \left( T_{\tau\tau} + T_{\sigma\sigma} \right) \right\rangle = \left( {1\over 2} a {\partial \over \partial a} \right) \sum_n e^{-\beta a E_n(a)}~,
\ee
from which we see that:
\be
\langle n|T_{\sigma\sigma} |n\rangle = - {a^2\over \pi} {\partial E_n \over \partial a}. 
\ee
From the definition of the momentum, 
\eq{\label{momentum_def}
iP= \int_\Sigma dx\sqrt{g_{\Sigma}} \,\hat{n}^\mu \xi^\nu_\sigma T_{\mu\nu}=\int d\sigma \frac{a}{\sin\sigma}\frac{\sin\sigma}{a} \frac{1}{a} T_{\tau\sigma}=\frac{\pi}{a} T_{\tau \sigma}~,
}
we find the relation
\eq{
\langle n|T_{\tau\sigma} |n\rangle=\frac{a}{\pi}iP_n~.
}
This definition of momentum $P$ gives a conserved quantity only for particular choices of boundary conditions. 
Substituting these expressions into~\C{almostflow} gives
\eq{\label{boxedIB}
\boxed{{\partial E_n \over \partial \lambda}=-{1\over 2 \pi } \left( E_n {\partial E_n \over \partial a} + {1\over a} P_n^2 \right)
}~.
}

\subsection{Deformed Energy Spectrum}

Now we can solve the deformed inviscid Burgers' equation for the deformed energy levels. Note that this equation is nearly identical to the standard inviscid Burgers' equation for the $\TT$-deformed energy levels for a theory on the cylinder where the radius of the cylinder is exchanged with the $\AdS$ length scale. 

Therefore, we find that the deformed energy levels have the same solutions. For the case of states with $P_n=0$ in a conformal field theory, the deformed energy spectrum can be written explicitly
\eq{\label{energy_solution}
E_n=-\frac{\pi a}{\lambda}\left(1-\sqrt{1+\frac{2\lambda E_n^{(0)}}{\pi a}}\right)~.
}
If the ground state energy $E_0^{(0)}$ is negative then the deformed ground state energy (\ref{energy_solution}) become complex if $\lambda$ exceeds $\lambda_{\mathrm{max}} = \frac{a \pi}{2 | E_0^{(0)} | }$. A negative ground state energy does occur for $\AdS_2$ in global coordinates, as was seen in the discussion of the Weyl anomaly found around equation~\eqref{Lioustress}.%
\footnote{Using the thermodynamic Bethe ansatz, similar complex values of the ground state energy in flat space have been related to CDD ambiguities in the $S$-matrix of the deformed theory~\cite{Mussardo:1999aj}.}

The zero-point energy behaves differently for a CFT in $\AdS_2$ studied in Poincar\'e coordinates rather than global coordinates \cite{Spradlin:1999bn}. For example, in the Poincar\'e patch the mode expansion for a free scalar is continuous rather than discrete; morally speaking, a scalar in Poincar\'e coordinates behaves more like a theory on the plane rather than the cylinder, while the theory in global coordinates behaves like a theory on the cylinder (or the strip). In particular, the Casimir energy in Poincar\'e coordinates vanishes even though the vacuum state is identical to that in global $\AdS_2$ (while it is the same state, the Poincar\'e and global Hamiltonians are different operators, with different spectra). For the Poincar\'e Hamiltonian, since $E_0^{(0)} = 0$, the deformed ground state energy $E_0 ( \lambda ) = 0$ and the spectrum remains real for arbitrarily large values of $\lambda$.

Finally, we note that when $\lambda$ is small, we can treat the correction to the energy perturbatively, and find
\eq{\label{pertE}
E_n=\sum_m \lambda^m E_n^{(m)}= E_n^{(0)}-\lambda \frac{(E_n^{(0)})^2}{2 \pi a}+\lambda^2\frac{(E_n^{(0)})^3}{2 \pi^2 a^2}+\ldots~. }

\subsubsection{Example: Energy Correction for a Free Boson}

Let us check the solution of the inviscid Burgers' equation by computing the leading order in $\lambda$ correction to the energy for an example. 
Consider the free, real  boson on $\AdS_2$. Starting from the Lagrangian 
\be
\CL= \frac{1}{2} \partial_\mu \phi\partial^\mu \phi~,
\ee
the perturbation by the determinant of the stress energy tensor will be of the form in (\ref{leading_deformed_lagrangian}), namely
\be
\Delta \CL = - \frac{1}{4} \lambda (\partial_\mu \phi\partial^\mu \phi)^2~. 
\ee
We can expand $\phi$ in modes as
\be
\phi(x)=\sum_{n\in 2\IZ^+}\frac{1}{\sqrt{n\pi}}\left(a_n\cos(n \sigma) e^{i n\tau}+a_n^\dagger\cos(n\sigma) e^{-i n \tau}\right)~, 
\ee
where we have arbitrarily chosen Neumann boundary conditions.
The choice of boundary conditions will not affect the flow equation itself; rather it can affect the initial conditions for the flow equation by altering the undeformed spectrum. However, we note that this choice of boundary conditions implies that the total momentum in any energy eigenstate vanishes, as we show explicitly in Appendix \ref{app:perturbative} around equation (\ref{momentum_vanishes}). Therefore, we will set $P_n = 0$ in this section.

The creation and annihilation operators satisfy the usual algebra, the $a_n$ annihilate the ground state, and the Hilbert space is the Fock space generated by the $a_n^\dagger$.

Now we can compute the energies,
\begin{align}\label{FFenergies}\begin{split}
E_N&=\frac{1}{2a}\big\langle N\big|\int \d\sigma (-(\partial_\tau \phi)^2+(\partial_\sigma \phi)^2)\big|N\big\rangle~,\\[.2cm]
&=\sum_{n,m} \frac{\sqrt{n m}}{2\pi a}\int d\sigma  \cos((n-m)\sigma)\;\big\langle N\big|a_n a_m^\dagger e^{i(n-m)\tau}+ a_n^\dagger a_m e^{-i(n-m)\tau}\big|N\big\rangle~,\\
&=\frac{1}{2a}\times 2\left(N+\sum_{n=1}^\infty n\right)%\int \frac{\sin\sigma}{a}d\sigma\\
=\frac{1}{a}\left(N-\frac{1}{12}\right)~,
\end{split}\end{align}
where here 
\eq{
|N\rangle=\prod_i (a_{n_i}^\dagger)^{m_i}|0\rangle\quad, \quad N=\sum_{i}n_i m_i~.
}
The shift by $-1/12$ on the last line of~\eqref{FFenergies} amounts to the anomalous contributions $\theta_{++},\theta_{--}$ to the stress tensor from~\eqref{Lioustress}.
Similarly we can compute the perturbative correction from the $\TT$ deformation, 
\begin{align}
    %\begin{split}
        \Big\langle N\Big|& \int d\sigma \sqrt{g}\frac{\sin^4\sigma}{4a^4}\left((\partial_\tau\phi )^4+(\partial_\sigma \phi)^4- 2(\partial_\sigma\phi \partial_\tau \phi)^2\right)\Big|N\Big\rangle
        \nonumber\\
        &=\sum_{n_i}\frac{\sqrt{n_1n_2n_3n_4}}{4\pi^2}\nonumber\\
        &\quad\times\int d\sigma \frac{a^2}{\sin^2\sigma}\frac{\sin^4\sigma}{a^4}\left( \prod_i \cos(n_i \sigma)+ \prod_i \sin(n_i\sigma)- 2\prod_{\{i,j\}}\cos(n_i \sigma )\sin(n_j\sigma )\right)
        \nonumber\\&\quad\times
        \Big\langle N\Big| {\rm permutations~of~}a_{n_i}a_{n_j}a_{n_k}^\dagger a_{n_\ell}^\dagger e^{i(n_i+n_j-n_k-n_\ell)\tau} \Big|N\Big\rangle~,
        \nonumber\\[.2cm]
        &=\frac{1}{4\pi^2 a^2}\times4\times \frac{\pi}{8}(6+6-2)\Big(N+\sum_n n\Big)^2~,\\
        &=\frac{\left(N-\frac{1}{12}\right)^2}{2\pi a^2}=
        \frac{ (E^{(0)})^2}{2 \pi }~. \nonumber
    %\end{split}
\end{align}
Identifying the above expression with the first order correction to $-a\partial_\lambda E_N$, we find that 
\be\label{perturbative_energies_result}
E_n^{(1)}=-\frac{1}{2\pi a}(E_n^{(0)})^2~,
\ee
which matches the prediction from \eqref{pertE}.  See Appendix \ref{app:perturbative} for more details.

\subsection{High Energy Behavior}

Let us now consider the high energy behavior of the $\TT$-deformed theory. Recall we have assumed that the UV behavior of the  undeformed theory is described by a CFT. This implies that the high energy density of states has Cardy behavior
\be
\rho(E^{(0)})\sim \exp\Big[{\sqrt{\frac{c}{3}E^{(0)}
}}\Big]~. 
\ee
On the other hand the high energy behavior ($E_n \gg a/\lambda$) of the deformed energy scales  as 
\be
E_n= \sqrt{\frac{aE_n^{(0)}}{\lambda}} \quad \Rightarrow \quad E_n^{(0)}\sim \frac{\lambda E_n^2}{a}.
\ee
This implies that the high energy density of states in the deformed theory has the Hagedorn growth
\be
\rho(E_n)\sim \exp\Big[{\sqrt{\frac{c\lambda}{3a}} E_n}\Big]~,
\ee
characteristic of a string theory.

In \cite{Barbon:2020amo} it was shown that this high energy behavior generically leads to a negative specific heat for flat space $\TT$-deformed theories. Because our flow equation for the energies is identical to that for field theories on a flat cylinder, we expect that a similar conclusion applies to theories in $\AdS_2$ deformed by the $\TT$ operator. 

\section{Flow Equation for the Partition Function}\label{sec:flow_partition}

To close our discussion, we would like to derive a flow equation for the partition function from the energy formula~\C{boxedIB}. Flow equations for the partition functions of $\TT$-deformed theories on flat tori have been studied in~\cite{Cardy:2018sdv, Datta:2018thy, Aharony:2018bad}.  For simplicity, let us restrict to the zero momentum sector. Starting with \footnote{Note that here, as opposed to the analysis on the conformal cylinder, the partition function has a temperature $\frac{1}{\beta a}$ instead of $\frac{1}{\beta}$. The reason is that in making the choice $\xi^\mu_t={1 \over a}\left(\frac{\partial}{\partial \tau}\right)^\mu$, we have chosen the Hamiltonian operator to be dual to translations in $(a\tau)$ rather than just $\tau$. Thus, the corresponding temperature is appropriately multiplied by $\frac{1}{a}$ as compared to the case where energy is dual to $\tau$.}
\be
Z=\sum_n e^{-\beta a E_n}~,
\ee
we differentiate and apply~\C{boxedIB} to find:
\eq{\label{Zflow}
{\partial Z\over \partial \lambda}  & = \sum_n \left(-\beta a {\partial E_n \over \partial \lambda } \right) e^{-\beta a E_n}~, \\
&= \sum_n \left( \frac{\beta a}{2 \pi} E_n {\partial E_n \over \partial a} \right) e^{-\beta a E_n}~. 
}
We can replace the explicit $E_n {\partial E_n \over \partial a}$ by $\partial_\beta \partial_a$ and correct the resulting expression to agree with~\C{Zflow} finding, 
\be
{\partial Z\over \partial \lambda} = {1\over 2\pi a} \left\{ \partial_\beta\partial_a - {\beta\over a}\partial_\beta^2 - {1\over \beta} \partial_a \right\} Z~.
\ee
It would be interesting to derive this flow equation using the methods studied in~\cite{Cardy:2018sdv}.

\section*{Acknowledgements}

T.\,D.\,B. is supported by the Mafalda and Reinhard Oehme Postdoctoral Fellowship in the Enrico Fermi Institute at the University of Chicago.
C.\,F. and S.\,S. are supported in part by NSF Grant No. PHY1720480, and C.~F. acknowledges support from the divisional MS-PSD program at the University of Chicago.   E.\,J.\,M. is supported in part by DOE grant DE-SC0009924.

% ==============================================================================

 \newpage

\appendix

\section{Factorization of \texorpdfstring{$\TT$}{TT}}
\label{app:2point}

Consider the two point function 
\eq{
C(x,y) &:= (I^{ac'}I^{bd'}-g^{ab}g^{c'd'})T_{ab}(x)T_{c'd'}(y)~,\\
T_{ab}(x) &= e_{a\mu}(x) e_{b\nu}(x)T^{\mu\nu}(x)~,
}
where $I_{ab'}$ is the parallel transport of the frame bundle from $x\to y$. For the metric (\ref{ads_metric}), the spin connection components are given by
\begin{align} \label{spinconnection}
    \omega_\tau^{ab} = \begin{bmatrix} 0 & - \cot ( \sigma ) \\ \cot ( \sigma ) & 0 \end{bmatrix}~, \qquad
    \omega_\sigma^{ab} = 0~.
\end{align}
We will only be interested in two point functions in which $x,y$ are spatially separated. This is critical because along this direction, the spin connection $\omega_\sigma^{ab}$ is trivial and hence $I_{ab'}=\delta_{ab'}$. 

As discussed in \cite{Allen:1985wd}, the tensor structure of the two point function 
\eq{\label{2pointT}
\big\langle T_{ab}(x) T_{c'd'}(y)\big\rangle~, 
}
is only dependent on the frame bundle metric $\delta_{ab}$, the parallel transport matrix $I_{ab'}$, and the unit normalized vectors
\eq{
n_\mu=\partial_\mu^{(x)} d(x,y)~,\quad \quad m_\nu=\partial_\nu^{(y)}d(x,y)~,
}
where $d(x,y)$ is the geodesic distance between $x,y$. Using the fact that $n_\mu,m_\nu$ satisfy \eq{\label{paralleltrans}
n_a+I_a^{~b'}m_{b'}=0~,
}
where $n_a=e_a^\mu n_\mu$, and similarly for $m$, we can eliminate all dependence in the tensor structure on $m_\nu$. Additionally, since we are taking $x,y$ spatially separated, $I_{ab'}=\delta_{ab'}$. Therefore, we find that all of the tensor structure can be reduced to the ``trivial" frame bundle metric $\delta_{ab}$, the trivial parallel transport tensor $\delta_{ab'}$, and the unit vector $n_a$. 

Since the two point function \eqref{2pointT} is a coordinate scalar and the spin connection is trivial in the direction of transport, the covariant derivative is simply the partial derivative. Additionally, the derivative of the scalar function reduces to the derivative of the 2-point function \eqref{2pointT} with indices contracted. 

Let us introduce the tensor 
\eq{
\epsilon_{ab'}=\epsilon_{ac}I^c_{~b'}~,
}
where $\epsilon_{ab}$ is the standard alternating tensor of two indices. This tensor obeys the property
\eq{
\epsilon_{ab'}\frac{\partial}{\partial x^\mu} =\epsilon_{\mu b'}\frac{\partial}{\partial x^a}+\epsilon_{\mu a}I^c_{~b'}\frac{\partial}{\partial x^c}~,
}
where here we mean $\frac{\partial}{\partial x^a}=e^\mu_{~a}\frac{\partial}{\partial x^\mu}$. Then using the fact that one can rewrite 
\eq{
C(x,y)=\epsilon_{ac'}\epsilon_{bd'}\langle T^{ab}(x) T^{c'd'}(y)\rangle~,
}
we can compute 
\eq{
\frac{\partial}{\partial x^e} C(x,y)&=\frac{\partial}{\partial x^e} \epsilon_{ac'}\epsilon_{bd'}\langle T^{ab}(x) T^{c'd'}(y)\rangle\\
&=\epsilon_{ec'}\epsilon_{bd'}\frac{\partial}{\partial x^a}\langle T^{ab}(x)T^{c'd'}(y)\rangle+\epsilon_{ea}\epsilon_{bd'}I^{f}_{~c'}\frac{\partial}{\partial x^f}\langle T^{ab}(x)T^{c'd'}(y)\rangle\\
&=\epsilon_{ea}\epsilon_{bd'}I^{f}_{~c'}\frac{\partial}{\partial x^f}\langle T^{ab}(x)T^{c'd'}(y)\rangle\\
&=-\epsilon_{ea}\epsilon_{bd'}\frac{\partial}{\partial y^{c'}}\langle T^{ab}(x)T^{c'd'}(y)\rangle=0~,
}
where in going from the penultimate to final line we used the identity \eqref{paralleltrans} in conjunction with the fact that the 2-point function is a function of $d(x,y)$ and has effectively trivial tensor structure. 

Therefore we find that %$C(x,y)$
\eq{
C(x,y)=C\big(d(x,y)\big)=C_0~,
}
is independent of geodesic distance between $x,y$.
Since $C(x,y)=C_0$, it obeys
\eq{
\lim_{x\to y}C(x,y)=\lim_{x\to \infty}C(x,y)~. 
}
Using the fact that $\AdS_2$ is non-compact and two point functions obey clustering, it is clear that 
\eq{
C(x,y)=\epsilon_{ac'}\epsilon_{bd'}\langle T^{ab}(x)\rangle\langle T^{c'd'}(y)\rangle=C_0~,
}
and in particular 
\eq{
\TT(y):=\lim_{x\to y}C(x,y)=\lim_{x\to y}\epsilon_{ac'}\epsilon_{bd'}\langle T^{ab}(x)\rangle\langle T^{c'd'}(y)\rangle~.
}
Therefore, we find that the operator $\TT(x)$ factorizes. 

\newpage

\section{Perturbative Calculation of Energies}\label{app:perturbative}

In this Appendix, we will explicitly verify the perturbative result (\ref{perturbative_energies_result}) in an arbitrary energy eigenstate, which matches the prediction from the inviscid Burgers' equation to leading order in $\lambda$.

As in the main body of the paper, we work with a massless scalar $\phi$ in $\AdS_2$ which has the mode expansion
\begin{align}\label{phi_mode}
    \phi ( x ) = \sum_{n \in 2 \mathbb{Z}_+} \frac{1}{\sqrt{n \pi}} \left( a_{n} \cos ( n \sigma ) e^{i n \tau} + a_n^\dagger \cos ( n \sigma ) e^{- i n \tau} \right) , 
\end{align}
with the usual algebra $[ a_n, a_m^\dagger ] = \delta_{n m}$. The expansion (\ref{phi_mode}) is normalized so that the coefficient functions $u_n = \frac{1}{\sqrt{n \pi} } \cos ( n \sigma ) e^{i n \tau}$ satisfy
\begin{align}
    \langle u_n \mid u_m \rangle = \delta_{nm} ,
\end{align}
where $\langle \cdot \mid \cdot \rangle$ is the Klein-Gordon norm,
\begin{align}
    \langle \phi \mid \psi \rangle &= - i \int_{\Sigma} \, dx \, \sqrt{h} \, n^\mu \left( \phi^\ast \partial_\mu \psi - \psi \partial_\mu \phi^\ast \right)~, \nonumber \\
    &= - i \int_0^\pi \, d \sigma \, \left( \phi^\ast \partial_\tau \psi - \psi \partial_\tau \phi^\ast \right) ~.
\end{align}
Here $\sqrt{h} = \frac{a}{\sin ( \sigma )}$ is the induced metric on the spatial slice $\Sigma$ and $n^\mu = \frac{\sin ( \sigma )}{a} \left( \frac{\partial}{\partial \tau} \right)^\mu$ is the unit timelike normal to the surface.

Now consider an arbitrary energy eigenstate $| J \rangle$, where $ J = \{ j_2, j_4, \cdots \}$ is a multi-index. Explicitly,
\begin{align}
    | J \rangle = \sum_{m \in 2 \mathbb{Z}_+} \left( a^\dagger_m \right)^{j_m} \, | 0 \rangle~.
\end{align}
 We first note that, with the choice of Neumann boundary conditions in the mode expansion (\ref{phi_mode}), the total momentum in any energy eigenstate vanishes. To see this, we compute $P$ using the definition (\ref{momentum_def}), finding
\begin{align}\label{momentum_vanishes}
    \langle J \mid P \mid J \rangle &= - \int_0^\pi \, d \sigma \, \sum_{n, m} \frac{ \sqrt{nm} }{\pi} \sin ( n \sigma ) \cos ( m \sigma ) \Big\langle J \, \Big\vert \,  \left( a_{n} e^{i n \tau} + a_n^\dagger e^{- i n \tau} \right) \times \nonumber \\
    &\hspace{100pt} \left( a_{m} e^{i m \tau} - a_m^\dagger  e^{- i m \tau} \right) \, \Big\vert \, J \Big\rangle~, \nonumber \\
    &= - \int_0^\pi \, d \sigma \, \sum_{n, m} \frac{ \sqrt{nm} }{\pi} \sin ( n \sigma ) \cos ( m \sigma ) \left( j_m j_n \delta_{n, m} - ( j_m + 1 ) ( j_m + 1 ) \delta_{n, m} \right)~, \nonumber \\
    &= \int_0^\pi \, d \sigma \, \sum_n \frac{n}{\pi} \sin ( n \sigma ) \cos ( n \sigma ) \left( 1 + 2 j_n \right) .
\end{align}
However for any $n \in \mathbb{Z}$, $\int_0^\pi \sin ( n \sigma ) \cos ( n \sigma ) = 0$. Therefore the total momentum in any state $| J \rangle$ is zero.

Next we compute the energy $E_J$, which is defined in terms of the stress tensor by
\begin{align}
    E_J = \int \, n^\mu \xi^\nu \langle J \mid T_{\mu \nu} \mid J \rangle \, d \Sigma~,
    \label{energy_defn}
\end{align}
where the integral is taken over a constant $\tau$ slice, $n^\mu$ is a unit normal, and $\xi^\nu = \frac{1}{a} \left( \frac{\partial}{\partial \tau} \right)^\nu$. Using the $\AdS_2$ metric (\ref{ads_metric}), this gives
\begin{align}
    E_J &= \frac{1}{a} \int_0^\pi \sqrt{g_{\sigma \sigma}} \, d \sigma \, \frac{\sin ( \sigma ) }{a} \langle J  \mid T_{\tau \tau} \mid J \rangle~, \nonumber \\
    &= \frac{1}{a} \int_0^\pi \, d \sigma \, \langle J \mid T_{\tau \tau} \mid J \rangle .
    \label{energy_integral}
\end{align}
The stress tensor for a massless scalar with background metric $g_{\mu \nu}$ is
\begin{align}
    T_{\mu \nu} = \partial_\mu \phi \partial_\nu \phi - \frac{1}{2} g_{\mu \nu} g^{\rho \sigma} \partial_\rho \phi \partial_\sigma \phi ,
    \label{massless_scalar_stress_tensor}
\end{align}
so in our case we have
\begin{align}
    T_{\tau \tau} = \frac{1}{2} \left( \frac{\partial \phi}{\partial \tau} \right)^2 + \frac{1}{2} \left( \frac{\partial \phi}{\partial \sigma} \right)^2 .
\end{align}
We must compute the expectation value of the squared $\phi$-derivatives in an arbitrary eigenstate. Using the mode expansion (\ref{phi_mode}), the first of these expectation values is
\begin{align}
    \Big\langle J \; \Big\vert \; \Big( \frac{\partial \phi}{\partial \sigma} \Big)^2 \; \Big\vert \; J \Big\rangle &= \sum_{k, m} \frac{\sqrt{k m}}{\pi} \, \sin ( k \sigma ) \, \sin ( m \sigma ) \, \langle J \mid \left( a_k e^{i k \tau} + a_k^\dagger e^{- i k \tau} \right) \cr 
    &\qquad  \times \left( a_m e^{i m \tau} + a_m^\dagger e^{- i m \tau} \right) \mid J \rangle~, \nonumber \\
    &= \sum_{k, m} \frac{\sqrt{km}}{\pi} \sin ( k \sigma ) \sin ( m \sigma ) \left( ( j_k + 1 ) \cdot \delta_{k, m} + j_k \cdot \delta_{k, m} \right)~, \nonumber \\
    &= \sum_{m} \frac{m}{\pi} \sin^2 ( m \sigma ) \cdot ( 2 j_m + 1 ) .
\end{align}
All sums run over even positive integers unless otherwise specified. By an almost identical calculation,
\begin{align}
    \Big\langle J \; \Big\vert \; \left( \frac{\partial \phi}{\partial \tau} \right)^2 \; \Big\vert \; J \Big\rangle = \sum_m \frac{m}{\pi} \cos^2 ( m \sigma ) \cdot ( 2 j_m + 1 ) .
\end{align}
Combining the two, we find
\begin{align}
    \Big\langle J \; \Big\vert \; \left( \frac{\partial \phi}{\partial \tau} \right)^2 + \left( \frac{\partial \phi}{\partial \sigma} \right)^2 \; \Big\vert \; J \Big\rangle = \sum_m \frac{m}{\pi} \cdot ( 2 j_m + 1 ) .
\end{align}
Therefore using (\ref{energy_integral}), we find
\begin{align}
    E_J &= \frac{1}{2a} \int_0^\pi \, d \sigma \, \bigg( \sum_{m \in 2 \mathbb{Z}_+} \frac{m}{\pi} \cdot ( 2 j_m + 1 ) \bigg) \nonumber \\[.2cm]
    &= - \frac{1}{12 a} + \frac{1}{a} \sum_{m \in 2 \mathbb{Z}_+} m \cdot j_m ,
    \label{undeformed_energy}
\end{align}
where we have regularized the sum as
\begin{align}
    \sum_{m \in 2 \mathbb{Z}_+} m = 2 \sum_{m=1}^{\infty} m = - \frac{1}{6} .
\end{align}

Next we compute the leading correction to the energy levels using the definition (\ref{TToperator}) of our perturbing operator, which enters the flow equation as
\begin{align}
    a \partial_\lambda E_J \, \big\vert_{\lambda = 0} &= - \int_0^\pi \, d \sigma \, \frac{\sin^2 ( \sigma ) }{a^2} \bigg( \frac{1}{4} \Big\langle J \; \Big\vert \; \left( \frac{\partial \phi}{\partial \tau} \right)^4 + \left( \frac{\partial \phi}{\partial \sigma} \right)^4 \; \Big\vert \; J \Big\rangle  \nonumber \\
    &\hspace{110pt}  - \frac{1}{2} \Big\langle J \; \Big\vert \; \left( \frac{\partial \phi}{\partial \tau} \right)^2 \, \left( \frac{\partial \phi}{\partial \sigma} \right)^2 \; \Big\vert \; J \Big\rangle \bigg) .
\end{align}
Again using the mode expansion (\ref{phi_mode}) and performing some straightforward oscillator algebra, the first two expectation values are
\begin{align}
    \Big\langle J \; \Big\vert \; \Big(  \frac{\partial \phi}{\partial \tau} \Big)^4 \; \Big\vert \; J \Big\rangle &= 3 \sum_{n} \frac{n^2}{\pi^2} \cos^4 ( n \sigma ) \cdot \left( 1 + 2 j_n + 2 j_n^2 \right) \nonumber \\
    &\qquad + 3 \sum_{n \neq m} \frac{n m}{\pi^2} \cos^2 ( n \sigma ) \cos^2 ( m \sigma ) ( 1 + 2 j_m + 2 j_n + 4 j_m j_n ) , \nonumber \\
    \Big\langle J \; \Big\vert \; \left(  \frac{\partial \phi}{\partial \sigma} \right)^4 \; \Big\vert \; J \Big\rangle &= 3 \sum_{n} \frac{n^2}{\pi^2} \sin^4 ( n \sigma ) \cdot \left( 1 + 2 j_n + 2 j_n^2 \right) \nonumber \\
    &\qquad + 3 \sum_{n \neq m} \frac{n m}{\pi^2} \sin^2 ( n \sigma ) \sin^2 ( m \sigma ) ( 1 + 2 j_m + 2 j_n + 4 j_m j_n ) .
\end{align}
Integrating over $\sigma$ then yields
\begin{align}
    \int \, d \sigma \, \frac{\sin^2 ( \sigma )}{a^2} \,  \Big\langle J \; \Big\vert \; \left(  \frac{\partial \phi}{\partial \sigma} \right)^4 \; \Big\vert \; J \Big\rangle &= \frac{9}{16 \pi a^2} \sum_{n} n^2 \cdot \left( 1 + 2 j_n + 2 j_n^2 \right) \nonumber \\
    &\qquad + \frac{3}{8 \pi a^2} \sum_{n \neq m} n m ( 1 + 2 j_m + 2 j_n + 4 j_m j_n ) ,
\end{align}
and
\begin{align}
    \int \, d \sigma \, \frac{\sin^2 ( \sigma )}{a^2} \, \Big\langle J \; \Big\vert \; \left(  \frac{\partial \phi}{\partial \tau} \right)^4 \; \Big\vert \; J \Big\rangle &= \frac{9}{16 \pi a^2} \sum_{n} n^2 \cdot \left( 1 + 2 j_n + 2 j_n^2 \right) \nonumber \\
    &\qquad + \frac{3}{8 \pi a^2} \sum_{n \neq m} n m \, ( 1 + 2 j_m + 2 j_n + 4 j_m j_n ) .
\end{align}
Similarly, for the mixed expectation value one finds
\begin{align}
    \int_0^\pi \, d \sigma \, \frac{\sin^2 ( \sigma )}{a^2} \Big\langle J \; \Big\vert \; \left( \frac{\partial \phi}{\partial \sigma} \right)^2 \left( \frac{\partial \phi}{\partial \tau} \right)^2 \; \Big\vert \; J \Big\rangle &= - \frac{1}{16 \pi a^2} \sum_n n^2 ( 1 - 2 j_n - 2 j_n^2 ) \nonumber \\
    &+ \frac{1}{8 \pi a^2} \sum_{n \neq m} n m ( 1 + 2 j_m ) ( 1 + 2 j_n ) .
\end{align}
Adding up the various contributions, this becomes
\begin{align}\label{perturbative_intermediate}
    a \partial_\lambda E_J \, \big\vert_{\lambda = 0} \, &= - \frac{1}{2 \pi a^2} \sum_{n} n^2 \cdot \left( j_n + j_n^2 \right) - \frac{1}{8 \pi a^2} \cdot \sum_{n \neq m} n m ( 1 + 2 j_m ) ( 1 + 2 j_n ) .
\end{align}
Here we have dropped terms of the form $\sum_{n \in 2 \mathbb{Z}_+} n^2$, which vanish upon performing zeta function regularization. 

The sums on the right hand side of (\ref{perturbative_intermediate}) can be combined and regularized in the same way that we have done before. Doing this, and identifying $\partial_\lambda E_J \, \big\vert_{\lambda = 0}$ as the leading correction $E_J^{(1)}$ to the energies, we find
\begin{align}\label{perturbative_appendix_final}
    E_J^{(1)} &= - \frac{1}{2 \pi a} \left( \frac{1}{144 a^2} - \frac{1}{6 a^2} \sum_{n \in 2 \mathbb{Z}_+} n j_n + \frac{1}{a^2} \sum_{n, m \in 2 \mathbb{Z}_+} n m j_n j_m \right)~, \nonumber \\
    &= - \frac{1}{2 \pi a} \left( E_J^{(0)} \right) ^2 .
\end{align}
Here $E_J^{(0)} = E_J$ is the undeformed energy computed in (\ref{undeformed_energy}). This reproduces (\ref{perturbative_energies_result}), as claimed.

\newpage
\bibliographystyle{utphys}
\bibliography{master}

\end{document}